\documentclass{llncs}
\usepackage{makeidx}
\usepackage{graphicx}
\begin{document}
%
%
\pagestyle{headings}  
\addtocmark{Hamiltonian Mechanics} 
\mainmatter              
\title{Does your Bug Tracking tool suit your needs?}
\subtitle{A Study on Open Source Bug Tracking tools}

\author{Sai Anirudh Karre\inst{1},  Anveshi Shukla\inst{2} \and Y. Raghu Reddy\inst{3}}
\institute{Software Engineering Research Center\\ International Institute of Information Technology, Hyderabad, India\\
\email{sai.anirudh@research.iiit.ac.in\\ anveshi.shukla@research.iiit.ac.in\\ raghu.reddy@iiit.ac.in}}

\maketitle 
\begin{abstract}
Bug tracking tools are vital for managing bugs in any open source as well as proprietary commercial projects. Considering the significance of using an appropriate bug tracking tool, we assess the features offered by 31 open source bug tracking tools and their significance of usage in open source projects. We have categorized these tools into different classes based on their features. We have also conducted a developer survey by working with open source software practitioners to understand the effectiveness of these tools in their day-to-day software development. We also explored StackOverFlow - a developer Q\&A forum to understand the developer experiences and challenges while using open source bug tracking tools. Our observations generated encouraging results that can used as a recommendation guide for open source software community to choose the best bug tracking tool based on their functional needs. Additionally, we have identified few features that are needed but not offered by most of these bug tracking tools.
\end{abstract}
\section{Introduction}
Open source software (OSS) development is a large team activity contributed by worldwide developer community. This community predominantly constitutes students and freelance software practitioners with various skill sets across the world. Almost all open source software teams depend on freeware tools to plan, code, test, track-report-fix bugs and market product(s). With huge end user client base, software production became easier with open source license \footnote{https://opensource.org/licenses}. For example, Apache HTTP Server project has huge client base as well as stroner developer base. It is the world's leading web server software since its origin in 1995. NetCraft 2015 Web Server Survey estimated that apache was serving 50.91\% of all active websites along with 49.19\% of top servers across all domains. This wide-spread popularity has triggered enthusiasm among the open source community to deliver more such products.

Open source initiatives do have regular challenges and competition from commercial and enterprise software development community. They lack structured execution on few crucial areas such as innovation, security, resource estimation, implementation, product support, inefficiency in documentation, lack of aggressive track of bugs and version control. For large projects, it is critical for any open source software manufacturer to manage the flood of bug requests while maintaining the existing fix repository. Recording and preserving bugs from end users across all versions is a big challenge for any open source software product. As a result, open source project owners look out for efficient bug tracking tools to handle these hurdles. Most of the successful open source products like Mozilla, Eclipse, Apache, Ubuntu and LINUX etc. depend on bug tracking tools to fix reported bugs and plan their new releases accordingly.

In most of the open source projects, choosing a tool requires consensus from all the open source project community. The team either favors most widely used tools or a tool on which one or more of the team members' have expertise on (synonymous to the "Golden Hammer" anti-pattern). In case of bug tracking tools, open source teams have to come out with similar consensus to decide on the best adoptable bug tracker that can meet their functional needs. However, not all bug tracking tools are good enough to address all the project needs due to their functional or nonfunctional properties. In such cases the open source team will have to either come up with their own customized bug tracker system or adopt the best possible tool. For example, The Debian - a UNIX like operating system uses its own internal email based bug tracking systems \cite{Francisco} to track and fix defects. However, choosing a bug tracking tool for an open source software project can be a quite interesting and troublesome task. This gap can be filled if there exists a criterion to evaluate the available bug tracking tools along with possible desired features. This can help the stake-owner(s) to choose the right open source bug tracking tool to meet their functional and business needs. Given this context, the major contributions of this paper are:

\begin{itemize}
\item Classify available open source bug tracking tools used by open source projects with features as criteria of evaluation.
\item A Comparative study on these tools and identification of most adopted features versus least adopted features by the open source software practitioners.
\item List down the features which are anticipated by software practitioners that are not offered by most of these bug tracking tools.
\end{itemize}

The primary objective our work is to help open source developer community to choose the right bug tracking tool for their new projects based on the size and scale of the product. We hope our evaluation of the bug tracking tools will help the stake-owner(s) to take right decisions while performing initial project plan for new open source product design and thereby helps them to improve their software quality and plan releases accordingly.

\section{Related Work}
McConnell was one of the initial proponents of using a bug tracking tool as a gauging tool for software release \cite{McConnell}. His methods helped software practitioners to decide when to release a product depending upon the quality of software. Ramsay \cite{Ramsay} introduced organized review process with defect tracking software so as to improve software quality. Fischer and Gall visualized bug tracker as a knowledge system \cite{Fischer} and captured various observations from different development communities. They have also analyzed how bug reports help on tracking evolution of product features \cite{Pinzger} using Bugzilla bug reporting system. McLaughlin has analyzed the generalized expectations and pitfalls of automated bug trackers \cite{McLaughlin} and shared his observations. Lintula and Koponen have explored maintenance process using defect management \cite{Lintula} as a criterion in four major open source software projects and shared their case studies. Francisco and Perez have studied the dependencies and effectiveness \cite{Francisco} of bug, bug notifications and communications between end users and developers in Debian's bug tracking tool. Premraj and Zimmermann have provided their recommendations and observations \cite{Just} along with improvements \cite{Sillito} for next generation bug tracking systems.

Guru et al. have performed an empirical analysis on four different bug trackers by comparing their features and listed out their drawbacks \cite{Abaee}.  Davies and Hanyu has performed an analysis on how the perspectives of bugs \cite{Davies} vary among bugs reported against each packages and have shared their observations. Jingyue and Stalhane are the first to perform an empirical analysis on process control of bug management \cite{Li} in bug tracking tools and have expressed their views on how this helps to improve software quality and assurance while manufacturing software. Yasufumi and Takeshi have proposed monitoring support for visualization \cite{Yasufumi} of bug trackers and have shared their observations. Yongsoo and Woosung have performed a comparative study on extraction methods of bug trackers \cite{Yongsoo}. Our work is more comprehensive compared to the works listed above and takes into consideration  24 attributes for the tools and 31 different bug tracking tools.d we shall try to derive additional information.

Several researchers have worked on improving and enhancing bug trackers for better software quality and effort estimation. Some researchers have also proposed methods to integrate the existing bug trackers to Integrated Development Environments (IDE) and project planners for efficient release management. To the best of our knowledge, there aren't any studies that try to understand the state-of-use of bug tracking tools from a large body of subjects. Moreover in this work, we identify new bug tracking features that do not yet exist in most bug tracking tools but if provided, programmers are willing to use. We also list some barriers to adoption.
\section{Study Setup}
As part of our analysis, we tried to answer the following research questions using an empirical study. The answers to these questions will help us in designing and implementing a recommendation system/tool in future. The tool can assist stake-owners with the selection of an optimal bug tracking tool based on their needs. Hence, as part of our work - we reached out to various open-source software communities and respective project managers to list down the bug tracking tools which are available and also used by them. The complete list of practically used open-source bug tracking tools are listed in table \ref{tools_list}. We considered only these tools as part of our work to perform our evaluation on their features. Based on the evaluation, we wanted to understand answers to below queries.

\begin{itemize}
\item{How easy is it to choose a bug tracking tool for a new open source project?}
\item{What do the open source practitioners perceive about the current bug tracking tools? Do these tools really meet all their requirements?}
\item{Check if the features desired by open-source developer community are supported in these current tools}
\end{itemize}

\begin{table}[h]
\caption{List of Bug Tracking tools}
\label{tools_list}
\centering
\begin{tabular} {|l|} \hline
\textit{Mantis, BugZilla, YouTrack, RedMine, }\\
\textit{ApacheBloodHound, JitterBug, GitHub,}\\ 
\textit{GoogleCode, RoundUp, BugNET, Savannah,}\\
\textit{Codeplex, teamatic, BugABoo, BitBucket,}\\
\textit{BugTraq, LaunchPad, RequestTracker, AceProject,}\\ 
\textit{JTrac, WebIssues, PhpBugTracker, Fossil,}\\ 
\textit{GNATS, BugAware, Trac, InformUp,} \\
\textit{eTraxis, Axosoft, Bugify, BUGTrack}\\ \hline
\end{tabular}
\end{table}

\subsection{Bug Tracking Tool Classification}
We have listed all the features offered by the tools listed in table \ref{tools_list}. We have compiled the list and have come up with a dataset\footnote{\url{https://app.box.com/s/bggs3iabqc8c1uame0qk7jift6fw06l9}} in this regard. This dataset consists of 23 features (which are listed as attributes in the dataset) captured individually for all the 31 bug tracking tools. The list of these attributes can be found in table \ref{attributes} based on their usage in open source community. The dataset contains a binary value assigned to each attribute for a given bug tracking tool. This binary value is based on its presence/absence of the features in a given bug tracking tool. For example, if $Doc$ feature is available in $Fossil$ bug tracking tool, '1' is assigned to this attribute in dataset, else '0' is assigned.

As the dataset is unlabeled with no class labels, we used unsupervised clustering methods to classify the data into clusters. The bug tracking tool dataset is too small, hence simple clustering techniques like K-Means \& Hierarchical clustering techniques \cite{Duda} were used to create class labels.

\noindent \textbf{K-Means Clustering:} Aims to estimate the unknown cluster centers (means) M = $\{\mu_1, \mu_2, \mu_3 \cdots, \mu_n\}$
based on the data points D =$\{x_1, x_2, x_3 \cdots, x_N\}$. Aims to minimize cost function where $\mu_i$ is the closest cluster center to $x_i$

\begin{equation}
J(M) = \sum\parallel x_i - \mu_i \parallel^2
\end{equation}

\noindent \textbf{Hierarchical clustering:} A cluster analysis that aims to build a hierarchy of clusters. All observations start in one cluster, which is iteratively split as one moves down the hierarchy. For large clusters this is slow where as for estimating small numbers of clusters it is efficient.

\subsection{Open-Source Community Survey}
Inputs from practitioners can go a long way in improving the features and solutions to common problems encountered while using bug tracking tools. Hence, we initiated an online survey consisting of open source software developers, testers, bug fixers, maintenance engineers, etc. to capture the user experience (from a functional perspective) while using bug tracking tools. Survey subjects were from Mozilla, DebainOS, Eclipse, Android, LibraOffice, OpenAFS, Drupal, OpenVPN, Ubuntu and Apache software foundation open-source communities. About 100 participants answered the questions provided in the survey. Questions pertaining to the current tool being used, its importance, features provided, its deficiencies, etc. were asked. Table \ref{survey} provides the details of our survey questions and response statistics captured.

\begin{table}[ht]
\centering
\caption{Attributes of Bug Tracker Dataset}
\label{attributes}
\begin{center}
\begin{tabular}{|p{1cm}|p{4cm}|p{6cm}|}\hline
\textbf{\textit{S.No}}&\textbf{\textit{Attribute Name}}&\textbf{\textit{About Attribute}}\\\hline
1&API&External API Support\\\hline
2&Testplan&Test Plan Integration\\\hline
3&Cwflow&Customizable Work Flow\\\hline
4&Cfields&Custom Field Support\\\hline
5&EmailNotif&Email Notification Support\\\hline
6&GUI&Customized GUI\\\hline
7&CLI&Command Line Support\\\hline
8&XMPP&XML support\\\hline
9&Comments&Comments for Bug Report\\\hline
10&Severity&Define Severity/Priority\\\hline
11&OpenId&Open Login Access\\\hline
12&LDAP/SHA1&Authentication Support\\\hline
13&RSS&RSS Feed Support\\\hline
14&OLA&Operation Level Agreement\\\hline
15&Report&Create Reports\\\hline
16&PM&Project Mgmt Support\\\hline
17&Doc&Document Generation\\\hline
18&Attachment&Attachment Support\\\hline
19&LinkDefect&Link/Clone/Merge bugs\\\hline
20&Emulate&Login as others (Admin feature)\\\hline
21&LocalLang& Localization Language Support\\\hline
22&SaaS/Hosted&SaaS/Hosted Deployment\\\hline
23&Road Map&Define Product Road Map\\\hline
24&Tool Name&Bug Tracker Name\\\hline
\end{tabular}
\end{center}
\end{table}

\begin{table}[h]
\centering
\caption{Survey Summary on Bug Trackers}
\begin{center} \label{survey}
\begin{tabular}{|p{5cm}|p{3cm}|p{3cm}|}\hline
\textbf{Questions}&\textbf{Responses}&\textbf{Response\%}\\ \hline
\multicolumn{3}{|p{7cm}|} {\textit{}}\\ \hline
\multicolumn{3}{|p{7.25cm}|}{\textit{Q1: How many years of work experience do you have as Software Practitioner (as a Bug reporter/ fixer/ developer)?}}\\ \hline
Less than 1 years &20&21\%\\ \hline
Between 1 and 5 years&33&34\%\\ \hline
More than 5 years&43&45\%\\ \hline
\multicolumn{3}{|p{7cm}|} {\textit{}}\\ \hline
\multicolumn{3}{|p{7cm}|}{\textit{Q2: Which Bug tracking tool is used for your Open Source Project?}}\\ \hline
Bugzilla&31&32\%\\ \hline
Apache Bloodhound&22&23\%\\ \hline
GitHub&14&15\%\\ \hline
Others&29&30\%\\ \hline
\multicolumn{3}{|p{7cm}|} {\textit{}}\\ \hline
\multicolumn{3}{|p{7cm}|}{\textit{Q3: How important the bug tracking tool is for your project?}}\\ \hline
High (We cannot work without it)&56&58\%\\ \hline
Moderate (We just need it)&28&29\%\\ \hline
Low (We don't use any)&12&13\%\\ \hline
\multicolumn{3}{|p{7cm}|} {\textit{}}\\ \hline
\multicolumn{3}{|p{7cm}|}{\textit{Q4: Does your bug tracking tool meet all your requirements?}}\\ \hline
High&45&47\%\\ \hline
Moderate&24&25\%\\ \hline
Low&27&28\%\\ \hline
\multicolumn{3}{|p{7cm}|} {\textit{}}\\ \hline
\multicolumn{3}{|p{7cm}|}{\textit{Q5: Does your bug tracking tool help you share all required details without any changes made to existing tool (with no additional API)?}}\\ \hline
High&36&37\%\\ \hline
Moderate&46&47\%\\ \hline
Low&15&15\%\\ \hline
\multicolumn{3}{|p{7cm}|} {\textit{}}\\ \hline
\multicolumn{3}{|p{7cm}|}{\textit{Q6: As a Bug fixer, Does your bug report explain everything you need to fix the issue?}}\\ \hline
High&28&29\%\\ \hline
Moderate&39&41\%\\ \hline
Low&29&30 \\ \hline
\multicolumn{3}{|p{7cm}|} {\textit{}}\\ \hline
\multicolumn{3}{|p{7cm}|}{\textit{Q7: Does your bug tracking tool provide all desired/standard features (Customizable Bug template/Reporting/Trend Analysis/Archive etc.)}}\\ \hline
High&56&58\%\\ \hline
Moderate&29&30\%\\ \hline
Low&11&11\%\\ \hline
\multicolumn{3}{|p{7cm}|} {\textit{}}\\ \hline
\multicolumn{3}{|p{7cm}|}{\textit{Q8: What are the features which are desirable but not provided by your bug tracking tool?}}\\ \hline
\multicolumn{3}{|p{7cm}|} {\textit{}}\\ \hline
\end{tabular}
\end{center}
\end{table}

\subsection{Developer Forum Analysis}
As part of day to day software development, software practitioners often use online forums to quickly debug and address issues. StackOverflow\footnote{\url{http://stackoverflow.com}} is one such popular online forum in the software development community. These forums generally contain questions and answers related to all types of problems faced by developers and also help them to understand how developers interact with each other to deduce a best solution supported by a voting process. In addition to the regular survey, analyzing forums like stack overflow can give us more qualitative end user response. This motivated us to conduct an analysis of bug tracking tools on StackOverflow form data dump. As a result, we could perform a comprehensive classification of flaws and desired features in bug tracking tools extracted from StakeOverflow form.

\section{Observations}
This section consists of our observations from bug tracking classification, Open-Source community Survey and Developer Forum Analysis.
\subsection{Clustering of Bug Tracking tools}
\begin{figure}
\centering \label{k}
\includegraphics [width=1\linewidth, height=9cm] {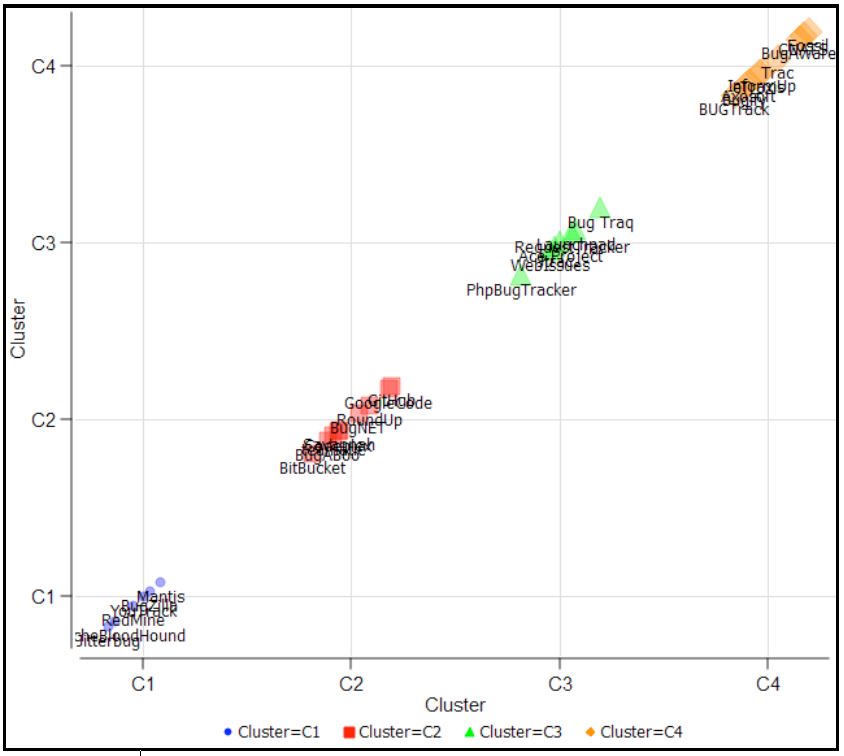}
\caption{K-Means Cluster on Bug Trackers}
\end{figure}

\begin{figure}
\centering \label{d}
\includegraphics [width=1\linewidth, height=9cm]{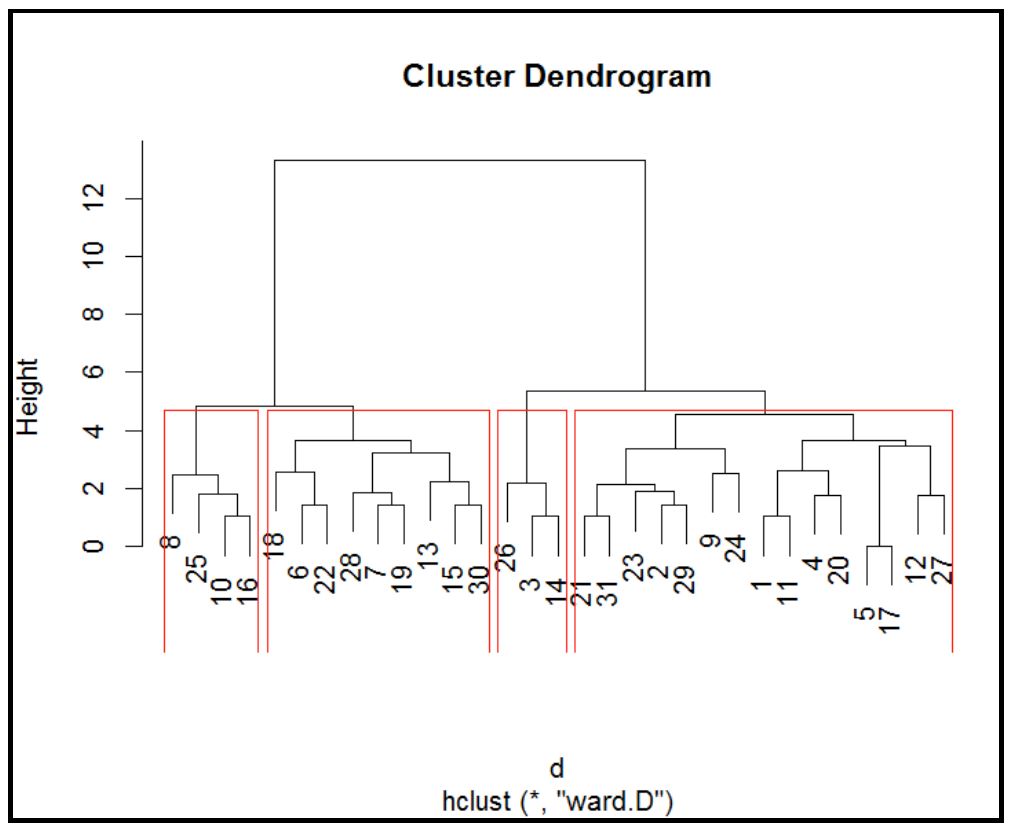}
\caption{Hierarchical Cluster on Bug Trackers}
\end{figure}

Using bug tracking tool dataset, we conducted unsupervised K-Means
and Hierarchical cluster algorithms to classify the bug tracking tools based on their features. We used R programming language\footnote{\url{https://www.r-project.org}} with Orangetext\footnote{\url{http://orange.biolab.si}} data mining tool for implementing the clustering. Figures \ref{k} and \ref{d} are one of the fold-outputs for cluster implementation. We clustered 31 bug tracking tools into 4 clusters and could intuitively observe figures \ref{k} that each cluster contains tools with almost similar features. Figure \ref{d} shows the dendrogram with tools classified into groups based on hierarchy they fall in.

For validation, we individually implemented these methods in 5 folds in turns and calculated Cosine Similarity among each fold. We used first fold $F_{0}$ as reference fold for other folds to calculate Cosine Similarity between both the methods for each fold. Cosine Similarity is a validation measure used to find the similarity between two vectors. 

Table \ref{validation} describes the validation results between every fold for K-Means clustering, Hierarchical clustering and between resultant vectors of K-Means - Hierarchical clusters. We could see that for every fold the cluster outputs are improved and are more similar. As we have small dataset, we were able to obtain 99\% similarity between the resultant vectors of both algorithms. Table \ref{avarage_clusters} shows the average cluster solution for our dataset of 31 bug tracking tools.

\textit{Cluster 1} contains tools that offer many features. Most of these tools provide features like test plan integration, customized work flow, custom fields, product road map planner, custom dashboards, user emulation and integration to external systems like project management and documentation management are offered. \textit{Cluster 2} contains simple tools with high support to code repository, localization and SaaS support. \textit{Cluster 3} gives good support to authentication features and standard reporting. \textit{Cluster 4} tools are strong in notifications and command line support. All these tools support very few features like comments, defining severity, file attachment support and customize GUI. It will be easy for Open source software developers who are working on new projects  at the clusters and plan for a right bug tracking tool with in a chosen cluster as per their functional need.

\begin{table}[ht]
\centering
\caption{Clustering Validation}
\label{validation}
\begin{center}
\begin{tabular}{|p{1cm}|p{2.10cm}|p{2.5cm}|p{1.5cm}|}\hline
\multicolumn{4}{|c|}{\textbf{Cosine Similarity Calculation}} \\
\hline
\textbf{\textit{Fold}}&\textbf{\textit{K-Means(A)}}&\textbf{\textit{Hierarchical(B)}}&\textbf{\textit{A and B}}\\\hline
$F_0 \& F_1$ & 0.9961 & 0.9976 & 0.99713\\\hline
$F_1 \& F_2$ & 0.9968 & 0.9979 & 0.99795\\\hline
$F_2 \& F_3$ & 0.9971 & 0.9981 & 0.98917\\\hline
$F_3 \& F_4$ & 0.9977 & 0.9988 & 0.9910\\\hline
$F_4 \& F_5$ & 0.9980 & 0.9991 & 0.9969\\\hline
\end{tabular}
\end{center}
\end{table}

\begin{table}[ht]
\centering
\caption{Average Clusters}
\label{avarage_clusters}
\begin{center}
\begin{tabular} {|c| p{5cm}|} \hline
\textbf{\textit{Cluster}}&\textbf{\textit{OSS Bug Tracking Tools}}\\ \hline
Cluster 1& \textit{Mantis, BugZilla, YouTrack, RedMine,
ApacheBloodHound, JitterBug}\\ \hline
Cluster 2& \textit{GitHub, GoogleCode, RoundUp, BugNET,
Savannah, Codeplex, teamatic, BugABoo, BitBucket}\\ \hline
Cluster 3& \textit{BugTraq, LaunchPad, RequestTracker, AceProject,
JTrac, WebIssues, PhpBugTracker}\\ \hline
Cluster 4& \textit{Fossil, GNATS, BugAware, Trac, InformUp,
eTraxis, Axosoft, Bugify, BUGTrack}\\ \hline
\end{tabular}
\end{center}
\end{table}

\subsection{Survey Discussion}
In this section we briefly summarize the discussions on the research questions posed in section 3 based on the results.
\noindent \textbf{Ease of choosing bug tracking tool}: There are many open source bug tracking tools with different features. Analyzing all of such tools will be a troublesome task. Hence, the major features of most used bug tracking tools were listed and categorized using machine learning techniques. The statistical analysis can help open source practitioners to choose a right tool that meeting their requirements.
\noindent \textbf{Efficiency of the bug tracking tool}: Based on the survey results we were able to deduce that 94\% of practitioners felt that bug tracking tools were critical to development. It was interesting to note that 88\% of users stated that their bug tracking tool had all the features they needed. However, a secondary look at the survey results from this segment revealed they had still listed down some features that they wish bug tracking tools had. 8\% of them felt that their needs were moderately met while the rest 4\% felt that they were habituated with tools that were not meeting their requirements. In spite of trying to align the team with specific tool usage, projects seem to face the issue of tools not meeting their requirements. Due to this, 12\% of users say that they depend on external API to support their bug tracking on top of existing tools. This becomes quite complicated to handle of request head from larger group of end users.
\noindent \textbf{Desired feature(s) not offered by existing tools:} From the survey we could clearly see that not all bug tracking tools provide anticipated features (even though they are widely used). Few of the features desired by open source teams are as follows:
\begin{itemize}
\item \textit{Knowledge Base Support:} All reported bugs may not lead to code change. In most cases, the cause for change be due
environment or by situation (due to data). In such cases if a work-around or a resolution were found, it would be really difficult to communicate the solution to all end users. If a Bug tracking tool is integrated or built with an in-house knowledge base system \cite{Ramsay}, it becomes easier for development teams to avoid duplicate bugs.
\item \textit{Release history:} As per Fischer, bug tracking tools with in-house collaborated module with revision history systems \cite{Fischer} help developers to generate automated release plan for the product. Hence, integration with version control system
would be an added advantage.
\item \textit{Defect Dependency:} In case of large-scale open source software development, it is really significant to have a metric to understand the dependency of a bug on all the available modules of a product. Incorporating defect dependency metric \cite{Karre} could address this issue.
\item \textit{Feature Tracker:} Unlike a project management system, it is really important to have a feature tracker in bug tracking tool to study the health of the feature introduced or removed from the product. Implementing a feature tracker \cite{Pinzger} will help developers link and track bugs relevant to specific feature.
\item \textit{Migration:} A bug tracker system should help developers to measure the condition of the project. The historical bug information is vital for any open source project. In case of projects where existing bug trackers are not helpful, new tools should be able to provide an option to migrate the data as dump so that it can be shared or reused by other collaborative systems.
\end{itemize}

\subsection{Discussion Forum Observations}
We studied Q \& A forum in StackeOverflow as another source to understand how open source community is looking for new bug tracking tools and below are most interesting observations:

\textit{Question: We have a startup and we're managing a bunch of code. Is there any open source integrated bug tracker and code review tool that we can install on the server and integrate with git? Tutorials to set this up would be wonderful.}

\textit{Question: So far, the choices seem overwhelming. I've looked at Mantis and Hiveminder. Unfuddle seems pretty close. I've avoided FogBugz for the price (and it seems like overkill) and Trac as I'm trying to avoid hosting something myself. Most of the existing solutions seem to be geared towards a team of developers and not for developer-client relations. Anyone have any recommendations?}

Above are couple of questions from many similar ones posted in
forum where open source developers were looking for an ideal
bug tracking tool to run their projects. Multiple users responded with list of tools with various options.

\textit{Question: What arguments might you use to support buying an existing bug tracking system? In particular, what features sound easy but turn out hard to implement, or are difficult and important but often overlooked?}

\textit{Question: Has anyone found a really good reason to need version control integration with the bug trackers?}

\textit{Question: Do you keep track of 'Potential Bugs' in your Bug Tracking System as well as the occurred ones? I mean, if you developed a piece of code that you realized in the end that under a specific condition it can fail but you don't have the time to implement a solution because of the strict deadlines}

Queries listed above are few of many other requests from software practitioners to understand the possibilities and scope of features supported by bug tracking tools. Few of the questions (not listed here) were about implementation of role hierarchy onto bug tracking tools. Similarly, there are many discussions on benefits of integrating code repositories, version control repositories, product management and test plan management with bug tracking tools.

As it can be seen from stack overflow forum, there are several questions pertaining to the features offered by the bug tracking tools and their usage context. Hence a comprehensive list of features, usage context and the corresponding classification will be really helpful to the practitioners. Our work can be enhanced with additional set of features, context information and additional tools to provide such recommendations.

\section{Limitations and Future Work}
In this paper we classified the most used open source bug tracking tools into clusters based on their features. The bug tracking tool classification is based on Unsupervised clustering methods. It is extremely difficult to evaluate the efficiency of results captured using unsupervised clustering algorithms. However we validated the results using Cosine Similarity and found the results to be accurate. Another limitation of our study is the size or the dataset and sample size of the survey. Given that the dataset covered all of the popularly used tools and communities, we feel that the results are significant and can be extrapolated.

We performed preliminary validations of our observations to
support our argument of choosing right bug tracking tool for a
given open source project. We reviewed the user experience by
analyzing discussions in StackOverFlow on features and issues
related to current bug tracking tools and have recorded their
feedback in this regard. Finally we proposed a few features that are most anticipated but are not provided by most of the bug tracking tools.

Our current research is oriented more towards to usage of bug tracking tools and recommending a specific bug tracking tool based on the needs of the stake-owners. An automated system for recommending bug trackers and providing alternate open source bug trackers for commercial bug trackers is in the pipeline. In future, we also plan to study the degree of improvement in software quality due to the usage of bug tracking tools.

\end{document}